\documentclass{article}%
\usepackage{amsfonts}
\usepackage{amsmath}
\usepackage{hyperref}
\usepackage{amssymb}
\usepackage{graphicx}%
\providecommand{\U}[1]{\protect\rule{.1in}{.1in}}
\begin{document}

\title{Image Charges Re-Imagined}
\author{H Alshal$^{1,2}$\thanks{{\footnotesize halshal@sci.cu.edu.eg}}, T
Curtright$^{2}$\thanks{{\footnotesize curtright@miami.edu}}, and S
Subedi$^{2}$\thanks{{\footnotesize sushil.subedi04@gmail.com}}\\$^{1}$Department of Physics, Cairo University, Giza, 12613, Egypt\\$^{2}$Department of Physics, University of Miami, Coral Gables, FL 33124-8046, USA}
\date{5 December 2018\\
\ \\
{\footnotesize \textquotedblleft Few things are harder to put up with than the
annoyance of a good example.\textquotedblright\ - Mark Twain}}
\maketitle

\begin{abstract}
We discuss the grounded, equipotential ellipse in two-dimensional
electrostatics to illustrate different ways of extending the domain of the
potential and placing image charges such that homogeneous boundary conditions
are satisfied. \ In particular, we compare and contrast the Kelvin and
Sommerfeld image methods.

\vfill%
\begin{center}
\includegraphics[scale=.475]{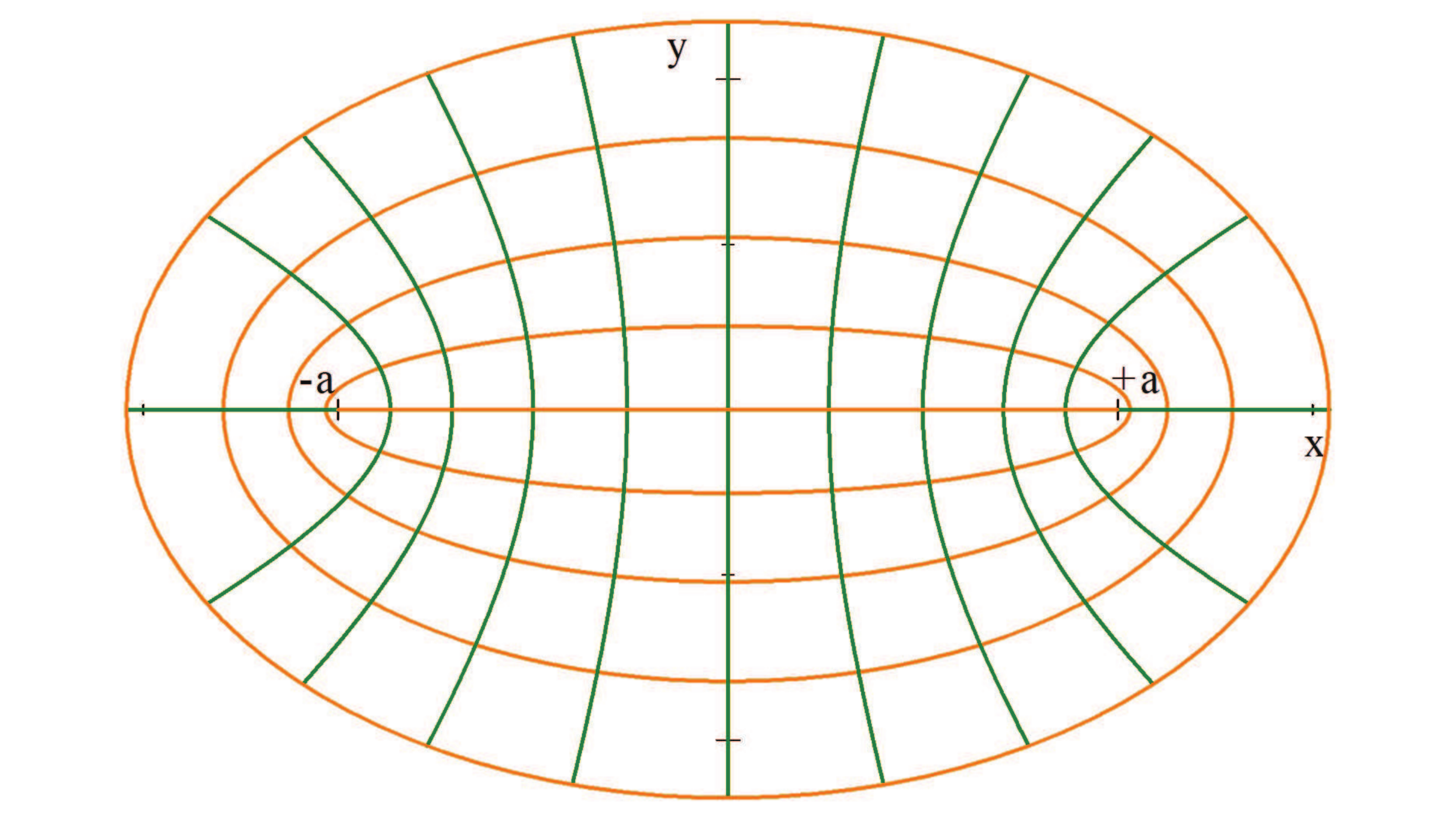}
\end{center}

\vfill

\end{abstract}

\newpage

\subsection*{Introduction}

When a source charge is placed near a real, grounded conductor, electrical
charge flows between the ground and the conductor. \ In the static limit, for
an idealized conductor, the resulting induced charge distribution is entirely
on the surface of the conductor. \ In this ideal static situation the
\emph{interior} of the conductor is an equipotential containing no charge, and
therefore not very interesting. \ However, for mathematical expediency, in
some cases one can easily \emph{imagine} a distribution of charge located
entirely \emph{inside} the conductor, instead of on the surface, which gives
exactly the same \emph{exterior} effects as the actual surface charge distribution.

All this is well-known, of course, but it may not be fully appreciated that
the imagined distribution of charge within the conductor is \emph{not}
uniquely determined.\footnote{In making this statement, we are not comparing
apples to oranges. \ It is well-known that different boundary conditions, such
as Dirichlet and Neumann, require different image charges. \ However, our
statement is correct even when we are dealing with only one set of boundary
conditions. \ In particular, we consider mixed homogeneous Dirichlet and
Neumann boundary conditions in this paper.} \ Perhaps the most interesting
aspect of this non-uniqueness lies in the mathematical freedom to choose the
interior of an idealized conductor (i.e. the domain of the image charge and
its potential) as an extension of the exterior (i.e. the domain of the real
source charge and its potential) to be almost \emph{any} imagined manifold,
with the only essential restriction being that the image and source domains
have in common a boundary, namely, the surface of the ideal
conductor.\footnote{The topology of the extended manifold may also be
re-imagined, but here we will not discuss that issue any further.}

This somewhat surprising mathematical freedom can be illustrated by a simple
example to be discussed below: \ The grounded two-dimensional (2D) ellipse.
\ Two image methods, established long ago by
\href{http://en.wikipedia.org/wiki/William_Thomson,_1st_Baron_Kelvin}{Thomson}
(a.k.a. Lord Kelvin) \cite{KelvinTait} and somewhat later by
\href{https://en.wikipedia.org/wiki/Arnold_Sommerfeld}{Sommerfeld}
\cite{Sommerfeld}, will be compared and contrasted. \ The image domains for
these two methods have different geometries, but nevertheless give exactly the
same physical results. \ The Kelvin method has the advantage that the
\href{https://en.wikipedia.org/wiki/George_Green_(mathematician)}{Green}
function \cite{Green} is usually easier to extend from the source domain to
the image domain. \ On the other hand, the Sommerfeld method has the advantage
that the location of the image is always obvious given the location of the
actual source charge. \ 

\subsection*{Kelvin versus Sommerfeld images --- a simple illustration}

The simplest example to illustrate the method of images is the problem of the
grounded plane, or rather, for the purposes of this paper, its 2D analogue,
the grounded line. \ The standard Green function on the entire plane follows
from the logarithmic potential, $-\frac{1}{2\pi}\ln\left(  \left\vert
\overrightarrow{r}\right\vert /R\right)  $, which involves an arbitrary scale
$R$. \ In terms of rectangular Cartesian coordinates on the entire plane,
$-\infty<x<+\infty$ and $-\infty<y<+\infty$, with orthogonal unit vectors,
$\widehat{x}$ and $\widehat{y}$, and with $\overrightarrow{r}=x~\widehat{x}%
+y~\widehat{y}$, the standard Green function is then%
\begin{equation}
g\left(  x_{1},y_{1};x_{2},y_{2}\right)  =-\frac{1}{4\pi}\ln\left(
\frac{\left(  x_{1}-x_{2}\right)  ^{2}+\left(  y_{1}-y_{2}\right)  ^{2}}%
{R^{2}}\right)  \ . \label{g}%
\end{equation}
Here $\left(  x_{1},y_{1}\right)  $ is the \textquotedblleft field
point\textquotedblright\ and $\left(  x_{2},y_{2}\right)  $\ is the
\textquotedblleft source point\textquotedblright. \ Note the symmetries
$g\left(  x_{1},y_{1};x_{2},y_{2}\right)  =g\left(  x_{2},y_{1};x_{1}%
,y_{2}\right)  =g\left(  x_{1},y_{2};x_{2},y_{1}\right)  =g\left(  x_{2}%
,y_{2};x_{1},y_{1}\right)  $. \ 

This $g$ is \href{https://en.wikipedia.org/wiki/Fundamental_solution}{a
fundamental solution} of the inhomogeneous equation%
\begin{equation}
\left(  \frac{\partial^{2}}{\partial x_{1}^{2}}+\frac{\partial^{2}}{\partial
y_{1}^{2}}\right)  g\left(  x_{1},y_{1};x_{2},y_{2}\right)  =-~\delta\left(
x_{1}-x_{2}\right)  ~\delta\left(  y_{1}-y_{2}\right)  \ , \label{gEqn}%
\end{equation}
with a 2D Dirac delta source on the right-hand side (RHS). \ This Green
function is therefore the logarithmic potential produced at the field point by
an ideal point charge located at the source point. \ For a general source
charge density on the plane, $\rho$, an electrostatic potential satisfying
\begin{equation}
\left(  \frac{\partial^{2}}{\partial x^{2}}+\frac{\partial^{2}}{\partial
y^{2}}\right)  \Phi\left(  x,y\right)  =-\rho\left(  x,y\right)  \ ,
\end{equation}
with some implicit boundary conditions, is then given by\
\begin{equation}
\Phi\left(  x,y\right)  =\int_{-\infty}^{+\infty}dX\int_{-\infty}^{+\infty
}dY~g\left(  x,y;X,Y\right)  ~\rho\left(  X,Y\right)  \ .
\end{equation}

To \textquotedblleft ground the line\textquotedblright\ $y=a$, and obtain the
Green function as well as a general potential on the half-plane $y>a$, such
that both satisfy homogeneous Dirichlet boundary conditions for $y=a$, it
suffices just to replace the Green function $g$ with the linear combination%
\begin{equation}
g_{o}\left(  x_{1},y_{1};x_{2},y_{2}\right)  =g\left(  x_{1},y_{1};x_{2}%
,y_{2}\right)  -g\left(  x_{1},2a-y_{1};x_{2},y_{2}\right)  \ .
\label{Kelvingo}%
\end{equation}
so that $g_{o}\left(  x_{1},a;x_{2},y_{2}\right)  =0$. \ This result has a
well-known interpretation, sometimes appropriately attributed to Kelvin, but
most often with no attribution at all. \ 

The interpretation follows from noting it is also true that
\begin{equation}
g_{o}\left(  x_{1},y_{1};x_{2},y_{2}\right)  =g\left(  x_{1},y_{1};x_{2}%
,y_{2}\right)  -g\left(  x_{1},y_{1};x_{2},2a-y_{2}\right)  \ .
\end{equation}
Thus the first term on the RHS is interpreted as the potential at the field
point $\left(  x_{1},y_{1}\right)  $ due to a point charge source at $\left(
x_{2},y_{2}\right)  $ while the second term is interpreted as the potential at
the field point due to a negative \textquotedblleft mirror
image\textquotedblright\ point charge source at $\left(  x_{2},2a-y_{2}%
\right)  $, the so-called Kelvin image. \ In this construction the half-plane
$y>a$ has been extended to the full plane, including all $y<a$, to allow
placement of the Kelvin image in the \textquotedblleft
unphysical\textquotedblright\ region below the grounded line. \ Consequently,
for all $y>a$ the equation (\ref{gEqn}) holds for $g_{o}$ as well as\ for $g$. \ 

The corresponding grounded potential for a general charge density situated in
the half-plane $y>a$ is then%
\begin{equation}
\Phi\left(  x,y\right)  =\int_{-\infty}^{+\infty}dX\int_{a}^{+\infty}%
dY~g_{o}\left(  x,y;X,Y\right)  ~\rho\left(  X,Y\right)  \ .
\end{equation}
From (\ref{Kelvingo}) it follows immediately that $\Phi\left(  x,a\right)
=0$. \ Moreover, the contributions to $\Phi$\ arising from the two terms in
$g_{o}$ may then be interpreted respectively as due to the real source density
$\rho\left(  X,Y\right)  $ above the grounded line, and an image source
density $-\rho\left(  X,2a-Y\right)  $ below that line. \ 

However, there are other ways to visualize the image charges. \ For example,
the Euclidean plane may be \emph{folded} along the grounded line to obtain two
copies of the half-plane $y>a$, with the negative image charge now located on
the second copy of the half-plane at the \emph{same} position as the source
point, namely, $\left(  x_{2},y_{2}\right)  $. \ This technique of employing a
second copy of the physical space is due to Sommerfeld, following in the
footsteps of Riemann to construct a branched manifold. \ The real beauty of
Sommerfeld's technique, in principle, is that doubling the physical space
\emph{obviously} works to provide the location of the image charges for all
homogeneous boundary condition potential problems in any number of dimensions.
\ But let's not get ahead of ourselves.

As it happens, for this particularly simple example, there is essentially
\emph{no difference} in the two methods. \ Mostly this is just because the
intrinsic geometry of the folded plane is indistinguishable from that of the
unfolded plane. \ Nevertheless, it is instructive to exhibit analytically the
parameterization of the folded space to be able to express the Green function
in Sommerfeld's approach. \ Here this is easily done: \ Represent the original
half-plane by points $\left(  x,y\right)  =\left(  x,a+w\right)  $ for $w>0$
and the second copy of the half plane by points $\left(  x,y\right)  =\left(
x,a-w\right)  $ for $w<0$. \ That is to say, the branched, folded plane is
represented by the points $\left(  x,y\right)  =\left(  x,a+\left\vert
w\right\vert \right)  $ for $-\infty<w<+\infty$. \ It is then important to
understand that point charges placed at the same $x$ but at different values
of $w$ do \emph{not} coincide, even though they may have the same $\left\vert
w\right\vert $. \ Such points\ with\ different $w$ but the same $\left\vert
w\right\vert $ are on opposite branches of the folded, doubled space.

The Green function on both branches of the folded space is now given by%
\begin{equation}
g\left(  x_{1},w_{1};x_{2},w_{2}\right)  =-\frac{1}{4\pi}\ln\left(
\frac{\left(  x_{1}-x_{2}\right)  ^{2}+\left(  w_{1}-w_{2}\right)  ^{2}}%
{R^{2}}\right)  \ ,
\end{equation}
for $-\infty<x_{1,2}<+\infty$ and $-\infty<w_{1,2}<+\infty$, and it again
provides a fundamental solution of
\begin{equation}
\left(  \frac{\partial^{2}}{\partial x_{1}^{2}}+\frac{\partial^{2}}{\partial
w_{1}^{2}}\right)  g\left(  x_{1},w_{1};x_{2},w_{2}\right)  =-~\delta\left(
x_{1}-x_{2}\right)  ~\delta\left(  w_{1}-w_{2}\right)  \ .
\end{equation}
But indeed, for this simple example, this $g$ is \emph{exactly the same
expression} as the previous Green function on the unfolded plane. \ Similarly,
grounding the line at $y=a$ is now accomplished by the linear combination%
\begin{equation}
g_{o}\left(  x_{1},w_{1};x_{2},w_{2}\right)  =g\left(  x_{1},w_{1};x_{2}%
,w_{2}\right)  -g\left(  x_{1},-w_{1};x_{2},w_{2}\right)  \ ,
\end{equation}
Moreover, the potential\ on the half-plane $y>a$, for a general $\rho$
distributed on that same half-plane, with the line $y=a$ grounded, is now%
\begin{equation}
\Phi\left(  x,w\right)  =\int_{-\infty}^{+\infty}dX\int_{0}^{+\infty}%
dW~g_{o}\left(  x,w;X,W\right)  ~\rho\left(  X,W\right)  \ ,
\end{equation}
where the field point is $\left(  x,y\right)  =\left(  x,a+w\right)  $ for
$w>0$. \ The contributions arising from the two terms in $g_{o}$ may then be
interpreted respectively as due to the real source density $\rho\left(
X,W\right)  $ above the grounded line, and the image source density
$-\rho\left(  X,-W\right)  $ also above the grounded line, but \emph{on the
opposite branch of the folded plane}.\ 

We wish to emphasize that the grounded line example is unique in its
simplicity as a 2D image system, since other examples have very different
geometries for their Kelvin and Sommerfeld image domains. \ We consider next a
situation where the alternative geometries of the combined source and image
manifolds for the Kelvin and Sommerfeld approaches are not so simply related,
namely, the grounded 2D ellipse.

\subsection*{Green functions for a 2D ellipse}

This problem is nicely solved using complex analysis, as has been known since
the 19th century (e.g. see the literature cited in \cite{Duffy}). \ However,
here we use real variables in anticipation of higher dimensional situations.
\ (Two Appendices discuss connections between our choice of real variables and
those of the conventional complex plane.) \ In terms of real elliptic
coordinates for the $xy$-plane\footnote{The straight line segment connecting
the two elliptical foci on the $x$-axis at $\pm a$\ is covered twice using
real elliptic coordinates.} as shown in the title page Figure,%
\begin{equation}
x=a\cosh u\cos v\ ,\ \ \ y=a\sinh u\sin v\ ,\ \ \ 0\leq u\leq\infty
\ ,\ \ \ 0\leq v\leq2\pi\ . \label{EllipticCoords}%
\end{equation}
Remarkably, the standard method to construct the 2D Laplacian Green function
as sums of harmonic functions (e.g. see \cite{AlshalCurtright,CurtrightEtAl})
now leads to an unusual form for the result.
\begin{equation}
G\left(  u_{1},v_{1};u_{2},v_{2}\right)  =-\frac{1}{4\pi}\left\vert
u_{1}-u_{2}\right\vert -\frac{1}{4\pi}\ln\left(  1+e^{-2\left\vert u_{1}%
-u_{2}\right\vert }-2e^{-\left\vert u_{1}-u_{2}\right\vert }\cos\left(
v_{1}-v_{2}\right)  \right)  \ . \label{G}%
\end{equation}
Note that in addition to being $2\pi$-periodic\footnote{As a consequence of
this $2\pi$-periodicity, $G$ could also be interpreted as the potential for an
infinite line of uniformly spaced point charges on the $uv$-plane, i.e. on the
covering space for the $\left(  u,v\right)  $ cylinder defined by
(\ref{EllipticCoords}). \ In that case the $\delta\left(  v_{1}-v_{2}\right)
$ on the RHS of (\ref{GEqn}) would be a
\href{https://en.wikipedia.org/wiki/Dirac_comb}{Dirac comb}. \ However, here
we are interested in only one copy of the cylinder, so this interpretation is
not relevant to the problem at hand.} in each of the $v$s this Green function
also has the following symmetries similar to those for $g$ above: $\ G\left(
u_{1},v_{1};u_{2},v_{2}\right)  =G\left(  u_{2},v_{1};u_{1},v_{2}\right)
=G\left(  u_{1},v_{2};u_{2},v_{1}\right)  =G\left(  u_{2},v_{2};u_{1}%
,v_{1}\right)  $. \ By construction, $G$ is again a fundamental solution of
the equation\footnote{At first sight it may be surprising that (\ref{GEqn}) is
the equation to be solved, since the elliptic coordinates defined in
(\ref{EllipticCoords}) involve a non-trivial metric. \ However, the metric
dependence factors out of the invariant Laplacian expressed in terms of those
elliptic coordinates. \ Thus, the covariant equation for the Green function,
namely, $\frac{1}{\sqrt{g}}~\partial_{\mu}\left(  \sqrt{g}~g^{\mu\nu}%
\partial_{\nu}G\right)  =-\frac{1}{\sqrt{g}}~\delta\left(  u_{1}-u_{2}\right)
~\delta\left(  v_{1}-v_{2}\right)  $, simply reduces to (\ref{GEqn}).}
\begin{equation}
\left(  \frac{\partial^{2}}{\partial u_{1}^{2}}+\frac{\partial^{2}}{\partial
v_{1}^{2}}\right)  G\left(  u_{1},v_{1};u_{2},v_{2}\right)  =-~\delta\left(
u_{1}-u_{2}\right)  ~\delta\left(  v_{1}-v_{2}\right)  \ , \label{GEqn}%
\end{equation}
and it incorporates some implicit boundary conditions. \ For example, all $v$
dependence in $G$ is exponentially suppressed as either $u_{1}$ or $u_{2}$
become infinite, with the other $u$ fixed. \ 

It is interesting to compare (\ref{G}) to the more well-known form given in
(\ref{g}). \ This is easily done using the elementary identity%
\begin{align}
&  \left(  \cosh u_{1}\cos v_{1}-\cosh u_{2}\cos v_{2}\right)  ^{2}+\left(
\sinh u_{1}\sin v_{1}-\sinh u_{2}\sin v_{2}\right)  ^{2}\nonumber\\
&  =\left(  \cosh\left(  u_{1}-u_{2}\right)  -\cos\left(  v_{1}-v_{2}\right)
\right)  \left(  \cosh\left(  u_{1}+u_{2}\right)  -\cos\left(  v_{1}%
+v_{2}\right)  \right)  \ . \label{AnID}%
\end{align}
Upon converting $x_{1,2}$ and $y_{1,2}$ to the elliptic coordinates in
(\ref{EllipticCoords}), this identity gives%
\begin{align}
g\left(  x_{1},y_{1};x_{2},y_{2}\right)   &  =-\frac{1}{4\pi}\ln\left(
\frac{a^{2}}{R^{2}}\left(  \cosh u_{1}\cos v_{1}-\cosh u_{2}\cos v_{2}\right)
^{2}+\left(  \sinh u_{1}\sin v_{1}-\sinh u_{2}\sin v_{2}\right)  ^{2}\right)
\nonumber\\
&  =-\frac{1}{4\pi}\ln\left(  2\cosh\left(  u_{1}-u_{2}\right)  -2\cos\left(
v_{1}-v_{2}\right)  \right)  -\frac{1}{4\pi}\ln\left(  \frac{a^{2}}{2R^{2}%
}\left(  \cosh\left(  u_{1}+u_{2}\right)  -\cos\left(  v_{1}+v_{2}\right)
\right)  \right) \nonumber\\
&  =G\left(  u_{1},v_{1};u_{2},v_{2}\right)  -\frac{1}{4\pi}\ln\left(
\frac{a^{2}}{2R^{2}}\left(  \cosh\left(  u_{1}+u_{2}\right)  -\cos\left(
v_{1}+v_{2}\right)  \right)  \right)
\end{align}
Therefore, for $u_{1}+u_{2}\neq0$ and real $v_{1}+v_{2}$, the difference $g-G$
is a non-singular, harmonic function, as must be the case for two fundamental
solutions of (\ref{GEqn}).

\subsubsection*{The Kelvin image method}

Characterized generally, albeit rather vaguely, the Kelvin image method makes
use of both the interior \emph{and} the exterior of the ellipse, placing
source and image charges in opposite regions so as to satisfy boundary
conditions. \ In the elliptic coordinate frame, an obvious construction of a
Green function for a grounded ellipse is given by the linear combination%
\begin{equation}
G_{o}\left(  u_{1},v_{1};u_{2},v_{2}\right)  =G\left(  u_{1},v_{1};u_{2}%
,v_{2}\right)  -G\left(  u_{1},v_{1};2U-u_{2},v_{2}\right)  \label{KelvinGo}%
\end{equation}
where the grounded ellipse consists of points given by $\left(  U,v\right)  $
for a fixed $U$ and $0\leq v\leq2\pi$. \ By construction, $G_{o}\left(
u_{1},v_{1};U,v_{2}\right)  =0$ for all $v_{2}$. \ From the symmetry of $G$ it
is also true that $G_{o}\left(  U,v_{1};u_{2},v_{2}\right)  =0$ for all
$v_{1}$. \ Some contour plots of $G_{o}$ are given in Appendix C, for $U=1$
and some representative field points.

For a general distribution of source charge either inside or outside the
grounded ellipse, as given by $\rho\left(  u,v\right)  $, the solution of
\begin{equation}
\left(  \frac{\partial^{2}}{\partial u^{2}}+\frac{\partial^{2}}{\partial
v^{2}}\right)  \Phi\left(  u,v\right)  =-k~\rho\left(  u,v\right)
\end{equation}
is then reduced to the evaluation of an integral involving $G_{o}$ and $\rho$.
\ In particular, for field points and actual sources outside the grounded
ellipse, the electric potential is
\begin{equation}
\Phi\left(  u_{1},v_{1}\right)  =k\int_{U<u_{2}\leq\infty}\int_{0\leq
v_{2}\leq2\pi}G_{o}\left(  u_{1},v_{1};u_{2},v_{2}\right)  \rho\left(
u_{2},v_{2}\right)  du_{2}dv_{2}\ .
\end{equation}
Here we have introduced $k$ as a 2D analogue of the Coulomb constant.

The first $G$ in (\ref{KelvinGo}) is universally interpreted as the potential
at field point $\left(  u_{1},v_{1}\right)  $ produced by a positive unit
point source at location $\left(  u_{2},v_{2}\right)  $. \ The second $G$ in
(\ref{KelvinGo}) is similarly interpreted as the potential at field point
$\left(  u_{1},v_{1}\right)  $ produced by another point-like, but in this
case negative, \emph{Kelvin image} at location $\left(  2U-u_{2},v_{2}\right)
$. \ However, for the grounded ellipse construction in (\ref{KelvinGo}) there
are some interesting --- perhaps unexpected --- features.

For both field and source points inside the grounded ellipse, such that $0\leq
u_{1},u_{2}\leq U$, the Kelvin image is always outside that ellipse with
$U\leq2U-u_{2}\leq2U$, and therefore the image is \emph{never} located at
infinity\footnote{This differs from a grounded circle in 2D (or sphere in 3D)
where the image is located by inversion and \emph{can} move toward infinity as
the source moves toward the center of the circle (or sphere). \ The limit
where the ellipse becomes a circle\ of radius $R$ is achieved here by
$R=\lim_{a\rightarrow0}\left[  a\cosh U\right\vert _{U=\ln\left(  2R/a\right)
}=\lim_{a\rightarrow0}\left[  a\sinh U\right\vert _{U=\ln\left(  2R/a\right)
}$. \ In this limit, only one copy of $\mathbb{E}_{2}$ is sufficient to solve
either the interior or the exterior problem using the Kelvin method. \ See
\cite{CurtrightEtAl}\ for a thorough discussion of the grounded circular ring
in 2D, where the standard Kelvin method is compared to the Sommerfeld method
in considerable detail.} as long as both $a\neq0$ and $U\neq\infty$. \ That is
to say, to implement an interior Green function construction inside a grounded
ellipse at $u=U$, it suffices to use a single point-like Kelvin image that
lies between the confocal ellipses at $u=U$ and $u=2U$. \ As expected, the
image is outside the source domain defined by $0\leq u\leq U$. \ In any case,
only \emph{one} copy of the plane $\mathbb{E}_{2}$ is sufficient for the
construction of the interior Green function.

On the other hand, for field and source points outside the grounded ellipse,
such that $U\leq u_{1},u_{2}\leq\infty$, the Kelvin image is inside that
ellipse, with $0\leq2U-u_{2}\leq U$, only so long as the source is not too
distant from the grounded ellipse. \ That is to say, the interior of the
original grounded ellipse contains the image only for $u_{2}\leq2U$. \ But if
the source is more distant, with $u_{2}>2U$, the chosen Kelvin image of the
point source passes through the line connecting the two foci and moves onto a
\emph{second} copy of $\mathbb{E}_{2}$ as also defined by
(\ref{EllipticCoords}) except with negative $u$. \ Therefore, for the
point-like Kelvin image construction of the complete exterior Green function
as expressed in (\ref{KelvinGo}), \emph{two} copies of the real plane are
required: \ One for $u>0$ and another for $u<0$. \ Effectively, the two
elliptical foci on the $x$-axis at $x=\pm a$ are connected by a straight line
segment that acts as a branch line \textquotedblleft doorway\textquotedblright%
\ joining together these two copies of $\mathbb{E}_{2}$. \ 

So, to solve the exterior electric potential problem for a grounded ellipse,
when\ real coordinates are used and point-like Kelvin images are located in an
obvious way, a branched manifold is necessarily encountered. \ To put it
another way, the actual, real interior of a\ grounded 2D elliptical conductor
is insufficient to accommodate the location of a single point-like Kelvin
image for an exterior point source, when that source is far from the
conductor. \ More interior space is needed! \ All this is represented
graphically in Figure 1.%
\begin{center}
\includegraphics[scale=.85]{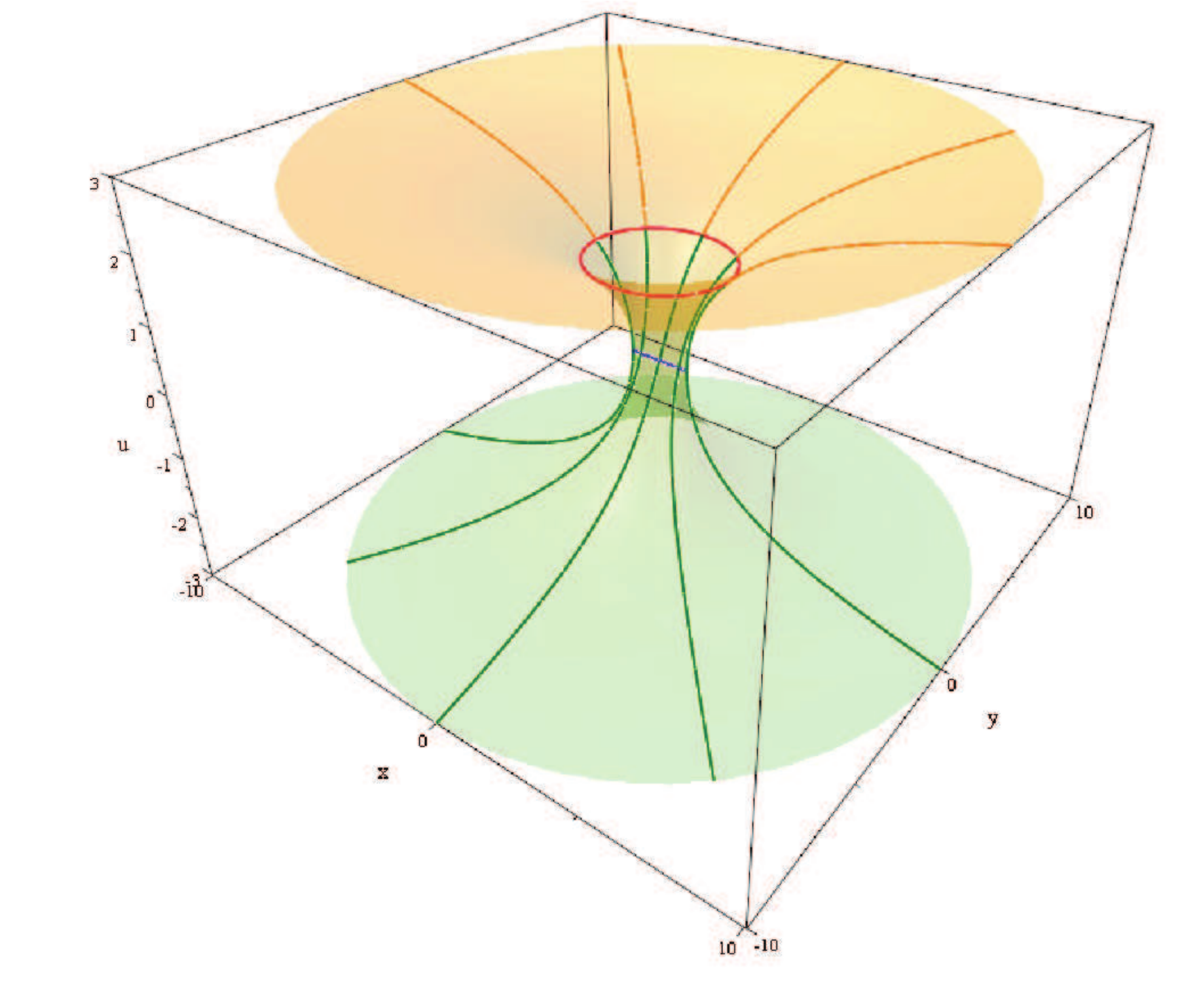}%
\\
Figure 1: \ Representative \textquotedblleft trajectories\textquotedblright%
\ for exterior sources (orange) and their Kelvin images (green) for a grounded
ellipse (red) with $U=3/2$. \ As a point source moves away from the red
ellipse along one of the orange curves, its image moves away from the red
ellipse along a corresponding (connected) green curve. \ A straight line
segment between the foci is shown in blue.
\end{center}

Another way to see these features for Kelvin images is through the use of
conformal mapping. \ By mapping a circle onto an ellipse, the standard Kelvin
image solution for a grounded circle is mapped onto an image solution for the
grounded ellipse. \ (Please see Appendix B.)

\subsubsection*{Mapping an infinite cylinder onto planes}

What is at work here is the fact that $G$ in (\ref{G}) is really a Green
function not just for the semi-infinite cylinder, with $u\geq0$, but actually
provides a solution to (\ref{GEqn}) for the infinite $uv$-cylinder, where
$-\infty\leq u\leq+\infty$, along with $0\leq v\leq2\pi$. \ So no matter where
the source is placed on that infinite cylinder, to construct $G_{o}$ such that
it vanishes at a fixed value of $u$, there is always room to accommodate a
Kelvin point image. \ 

The only open issue is then how to map the infinite $uv$-cylinder onto one or
more copies of the $xy$-plane. \ Sticking with the $x\left(  u,v\right)  $ and
$y\left(  u,v\right)  $ relations in (\ref{EllipticCoords}) gives a map that
produces two copies of $\mathbb{E}_{2}$\ as represented by the embedding shown
in Figure 1\footnote{For the chosen 3D embedding the surface has the
appearance of being intrinsically curved, but that is an artefact of the
parameterization. \ That part of the surface either above or below the blue
line in Figure 1 corresponds to an open subset of $\mathbb{E}_{2}$.} for the
case $U=3/2$. \ The original infinite $uv$-cylinder is flared out by the map
onto $x$ and $y$, both for large positive and for large negative $u$, but
pinched down to a straight line segment connecting the foci at $x=\pm a$ when
$u=0$, with that segment situated \textquotedblleft below\textquotedblright%
\ the grounded ellipse at $u=U$ ($=3/2$ in the Figure). \ This is the geometry
that underlies the Kelvin image method as applied here. \ The pinched line
segment has some obviously singular geometric features, but these are not pathological.

On the other hand, there is another clear choice to map the infinite
$uv$-cylinder onto planes that gives a different geometry. \ Rather than pinch
the cylinder shut in terms of $x$ and $y$, at $u=0$ or some other value of
$u$, the cylinder may be folded around the location of the grounded ellipse so
that the submanifold below the fold is just a \textquotedblleft mirror
image\textquotedblright\ of the submanifold above the fold. \ (Please see
Figure 2.\footnote{As in the previous Figure, for the chosen 3D embedding the
surface has the appearance of being intrinsically curved, but that is again an
artefact of the parameterization. \ That part of the surface either above or
below the red ellipse in Figure 2 corresponds to an open subset of
$\mathbb{E}_{2}$.}) \ This leads to the Sommerfeld image method which we
describe in detail in the following. \ The fold also has some obviously
singular geometric features, but again these are not pathological.

\subsubsection*{The Sommerfeld image method}

Consider the same exterior Green function situation using Sommerfeld images.
\ (The history of this alternate method is discussed in \cite{Eckert}.) \ In
this approach, the interior of the ellipse is eliminated, and two copies of
the plane outside the grounded ellipse are joined together along the grounded ellipse.

The new parameterization of both copies of the $xy$-plane outside the ellipse
with $u=U>0$, again written in terms of real elliptic coordinates,
is\footnote{If the apparent $du/dw$ slope discontinuity causes anxiety on the
part of the reader, one may take instead $u\left(  w\right)  =\left(
U^{2p}+w^{2p}\right)  ^{1/2p}$ for $p>1/2$, again with $-\infty\leq
w\leq\infty$. \ For example, see \cite{CurtrightEtAl}.\ \ However, the ensuing
complications in expressions involving Green functions are not worth making
this generalization, in our opinion.}%
\begin{align}
u  &  =U+\left\vert w\right\vert \ ,\\
x  &  =a\cosh\left(  U+\left\vert w\right\vert \right)  \cos
v\ ,\ \ \ y=a\sinh\left(  U+\left\vert w\right\vert \right)  \sin
v\ ,\ \ \ -\infty\leq w\leq\infty\ ,\ \ \ 0\leq v\leq2\pi\ .
\end{align}
So, when both field and source points are on the upper branch of the surface,
such that $0<w_{1},w_{2}<\infty$, the Green function is%
\begin{equation}
G\left(  w_{1},v_{1};w_{2},v_{2}\right)  =-\frac{1}{4\pi}\left\vert
w_{1}-w_{2}\right\vert -\frac{1}{4\pi}\ln\left(  1+e^{-2\left\vert w_{1}%
-w_{2}\right\vert }-2e^{-\left\vert w_{1}-w_{2}\right\vert }\cos\left(
v_{1}-v_{2}\right)  \right)  \ .
\end{equation}
But when the field point is on the upper branch and the source is on the lower
branch, albeit with the same convention $0<w_{1},w_{2}<\infty$, the Green
function is%
\begin{equation}
G\left(  w_{1},v_{1};-w_{2},v_{2}\right)  =-\frac{1}{4\pi}\left(  w_{1}%
+w_{2}\right)  -\frac{1}{4\pi}\ln\left(  1+e^{-2\left(  w_{1}+w_{2}\right)
}-2e^{-\left(  w_{1}+w_{2}\right)  }\cos\left(  v_{1}-v_{2}\right)  \right)
\ .
\end{equation}
In this approach the exterior Green function for a grounded ellipse is the
linear combination%
\begin{align}
G_{o}\left(  w_{1},v_{1};w_{2},v_{2}\right)   &  =G\left(  w_{1},v_{1}%
;w_{2},v_{2}\right)  -G\left(  w_{1},v_{1};-w_{2},v_{2}\right)
\label{SommerfeldGo}\\
&  =-\frac{1}{4\pi}\left\vert w_{1}-w_{2}\right\vert +\frac{1}{4\pi}\left(
w_{1}+w_{2}\right)  -\frac{1}{4\pi}\ln\left(  1+e^{-2\left\vert w_{1}%
-w_{2}\right\vert }-2e^{-\left\vert w_{1}-w_{2}\right\vert }\cos\left(
v_{1}-v_{2}\right)  \right) \nonumber\\
&  +\frac{1}{4\pi}\ln\left(  1+e^{-2\left(  w_{1}+w_{2}\right)  }-2e^{-\left(
w_{1}+w_{2}\right)  }\cos\left(  v_{1}-v_{2}\right)  \right)  \ ,\nonumber
\end{align}
assuming that both field and source points are on the upper branch, i.e.
$0\leq w_{1},w_{2}\leq\infty$. \ Otherwise, $G\left(  w_{1},v_{1};w_{2}%
,v_{2}\right)  =G\left(  w_{2},v_{2};w_{1},v_{1}\right)  $ and $G_{o}\left(
-w_{1},v_{1};w_{2},v_{2}\right)  =-G_{o}\left(  w_{1},v_{1};w_{2}%
,v_{2}\right)  $.

Remarkably, as the reader may readily verify, the expressions (\ref{KelvinGo})
and (\ref{SommerfeldGo}) give exactly the same functions on the $xy$-plane
when both field point $\left(  x_{1},y_{1}\right)  $ and source point $\left(
x_{2},y_{2}\right)  $ are located outside the grounded ellipse and on the
upper $\mathbb{E}_{2}$ branch, despite the differences in the Kelvin and
Sommerfeld image locations as evident upon comparing Figure 1 with the
following Figure 2.%

\begin{center}
\includegraphics[scale=.85]{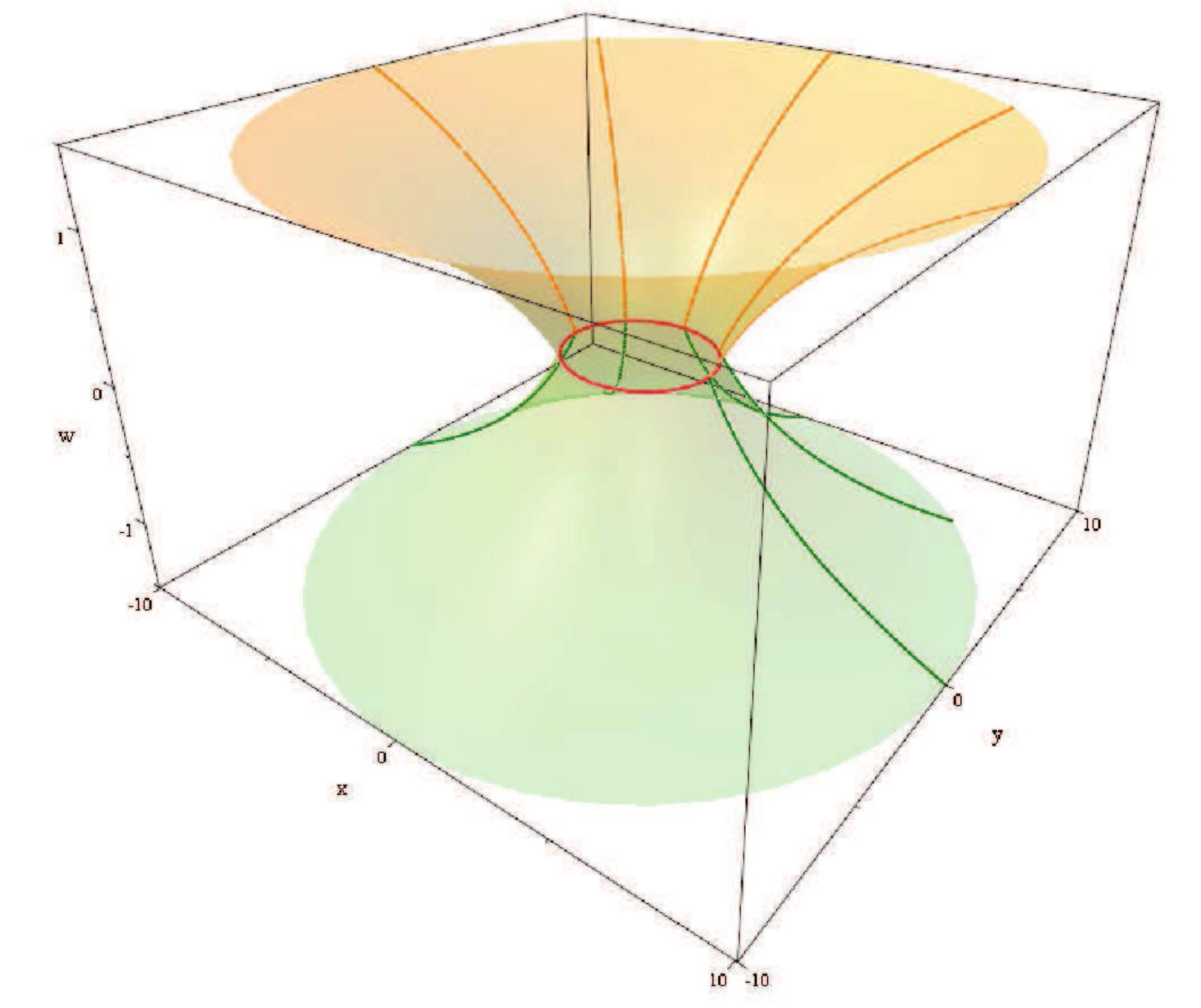}%
\\
Figure 2: \ Representative trajectories\ for exterior sources (orange) and
their Sommerfeld images (green) for a grounded ellipse (red), again with
$U=3/2$. \ All $\left(  x,y\right)  $ points inside the red ellipse are
excluded from the 2D manifold in this method.
\end{center}

Visualization of the features in these 3D Figures --- especially their
differences --- may be easier if 2D vertical slices are considered. \ In
Figure 3, the source and image domains along the $y$-axis are shown in green
for the Kelvin method and in orange for the Sommerfeld method. \ Particular
choices for point sources and their images are shown as small circles,
squares, or diamonds, for an ellipse whose $x=0$ points are shown in red.
\ The source domain is always the same --- namely, the planar region outside
the grounded ellipse --- no matter what image method is under consideration,
so the orange and green curves in the Figure are the same for $u>3/2$ or $w>0$.

In Figure 4, the source and image domains along the $x$-axis are shown in
green for the Kelvin method and in orange for the Sommerfeld method. \ As
before, particular choices for point sources and their images are shown as
small circles, squares, or diamonds, and the $y=0$ points on the ellipse are
shown in red. \ Once again, the source domain is always the same no matter
what image method is under consideration, but the image domains differ,
depending on how the manifold is extended beyond the source domain. \
\begin{center}
\includegraphics[scale=.5]{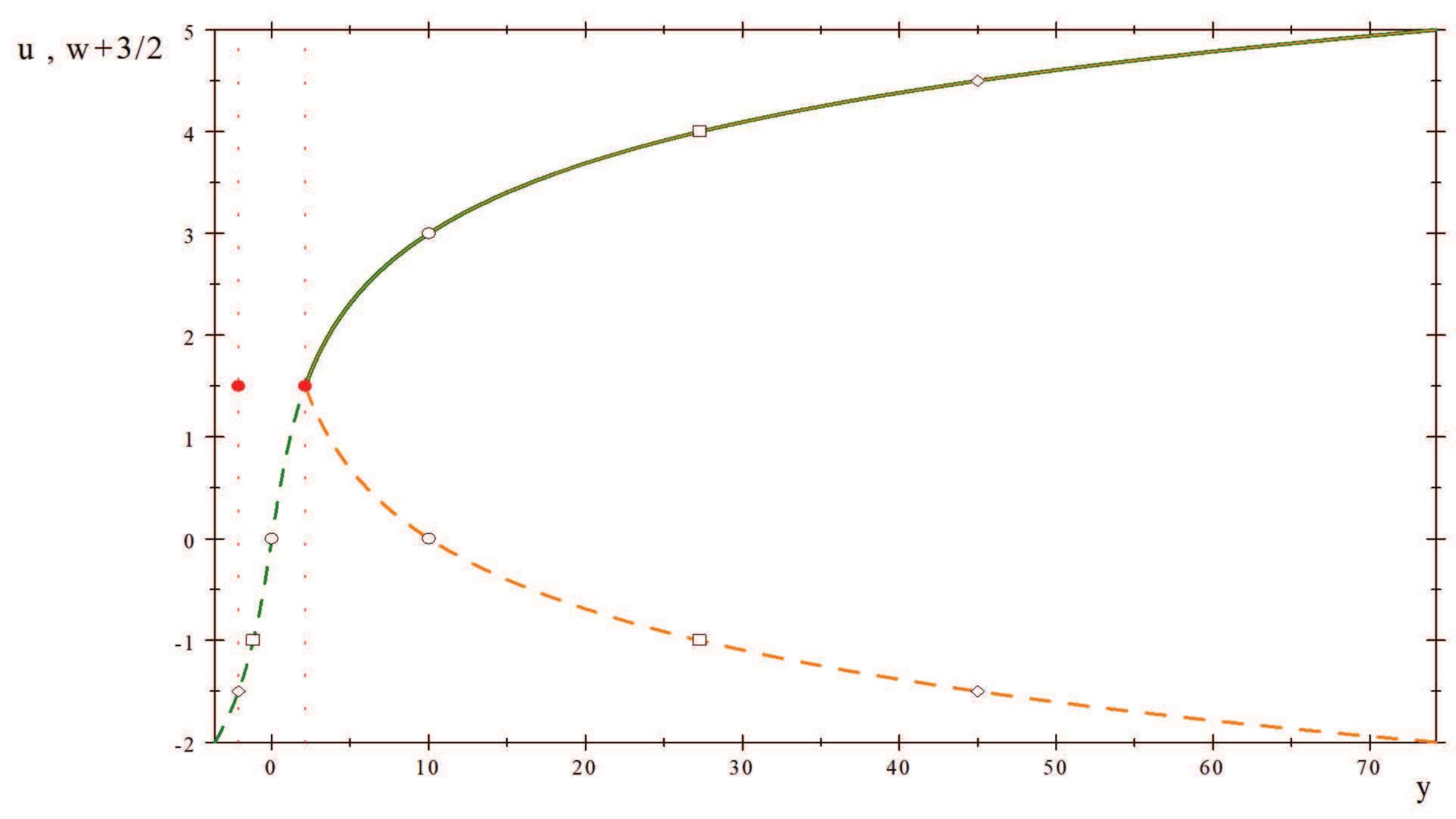}%
\\
Figure 3: \ Source and image domains for $x=0$, as solid and dashed curves,
respectively.
\end{center}
\begin{center}
\includegraphics[scale=.5]{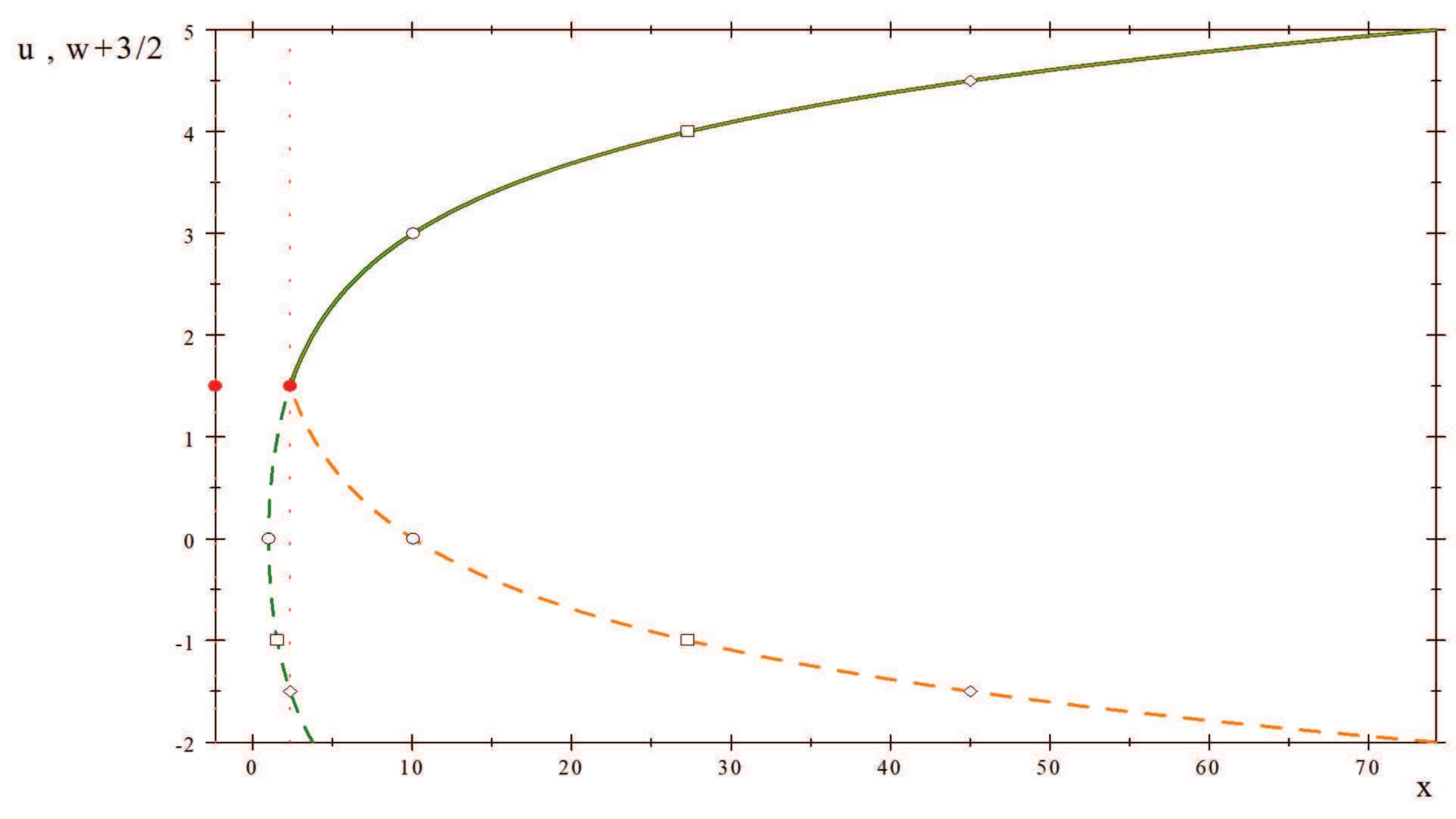}%
\\
Figure 4: \ Source and image domains for $y=0$, as solid and dashed curves,
respectively.
\end{center}
To summarize, it seems fair to say the image domain is largely determined just
by one's imagination.

\subsection*{Induced charge density}

The actual linear charge density induced on the grounded ellipse is
proportional to the normal component of the electric field evaluated in the
limit where the field point approaches the ellipse. \ It suffices to consider
the density induced by a unit point source outside the ellipse. \ Then the
relevant normal electric field is just $\left.  -\partial G_{o}/\partial
u_{1}\right\vert _{u_{1}=U}$ for the Kelvin image method, or $\left.
-\partial G_{o}/\partial w_{1}\right\vert _{w_{1}=0}$ for the Sommerfeld image
method. \ The results are the same, using either method. \ The situation of
interest for the external problem involves a unit source at $u_{2}>U$ or
$w_{2}>0$. \ 

In terms of the result for the Sommerfeld method, (\ref{SommerfeldGo}), we
find the linear charge density%
\begin{equation}
\lambda\left(  v_{1};w_{2},v_{2}\right)  =-\left.  \frac{\partial}{\partial
w_{1}}G_{o}\left(  w_{1},v_{1};w_{2},v_{2}\right)  \right\vert _{w_{1}%
=0,\ w_{2}>0}=\frac{1}{2\pi}\frac{e^{-2w_{2}}-1}{e^{-2w_{2}}-2e^{-w_{2}}%
\cos\left(  v_{1}-v_{2}\right)  +1}\ . \label{LinearDensity}%
\end{equation}
$\allowbreak$Note that the total charge induced by a $+1$ source is always
$-1$,
\begin{equation}
\int_{0}^{2\pi}\lambda\left(  v_{1};w_{2},v_{2}\right)  dv_{1}=-1\ ,
\end{equation}
\ even if the unit source is removed to infinity.\footnote{This is a
peculiarity of the long-range Coulomb potential in 2D --- it's logarithmic!
\ In 3D the charge induced on a grounded ellipsoid by a unit source outside
the sphere is not always $-1$, and in fact falls to zero as the source is
removed to infinity \cite{DassiosAgain,XueDeng}. \ For a grounded hyper-sphere
in $N$ spatial dimensions,\ it is an interesting exercise to show the induced
charge falls as a function of the source distance like $r^{2-N}$
\cite{AlshalCurtright}.} \ In that infinite limit, the induced charge density
becomes constant around the ellipse.%
\begin{equation}
\lambda\left(  v_{1};w_{2},v_{2}\right)  \underset{w_{2}\rightarrow
\infty}{\sim}-\frac{1}{2\pi}\ .
\end{equation}

Plots of the charge density for various selected source distances from the
grounded ellipse are straightforward to produce and evince all the expected
features when expressed in terms of our chosen elliptic coordinates.\vfill%
\begin{center}
\includegraphics[scale=.475]{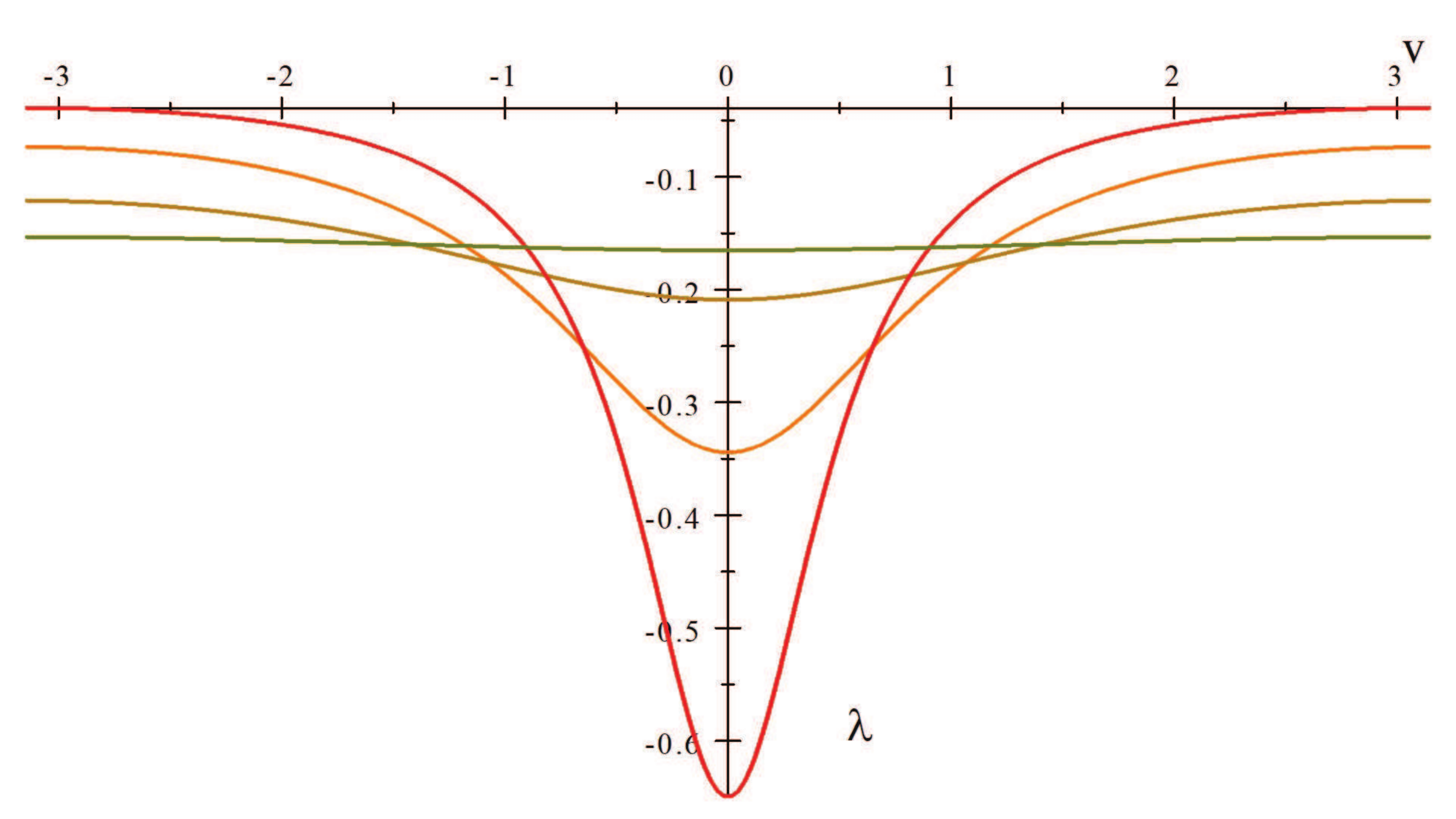}%
\\
Figure 5: $\ \lambda$ as a function of $v=v_{1}-v_{2}$ for various $w_{2}$.
\ Specifically,\ $w_{2}=1/2$ red, $w_{2}=1$ orange, $w_{2}=2$ sienna,
$w_{2}=4$ brown.
\end{center}

\vfill

\subsection*{A straight line limit}

A straight line limit of the ellipse is achieved by first setting $v=\pi/2$ in
(\ref{EllipticCoords}) so that $x\equiv0$, and then letting $a\rightarrow
\infty\ $and$\ u\rightarrow0$ so that $\lim_{a\rightarrow\infty,u\rightarrow
0}\left(  a\sinh u\right)  =y$ remains finite. \ The essential idea is that as
$a\rightarrow\infty$ the elliptical $\left(  u,v\right)  $ coordinates near
the center of the $x$-axis become just rectangular Cartesian coordinates,
$\left(  x,y\right)  $. \ This behavior is evident in the title page Figure,
even for finite $a$.

That is to say, let $u=y/a$ as $a\rightarrow\infty$ so that%
\begin{equation}
a\sinh u\rightarrow ay/a=y\ .
\end{equation}
At the same time, let $U=Y/a$ for $v=\pi/2$. \ Then $y\left(  U,\pi/2\right)
\rightarrow Y$ as $a\rightarrow\infty$. \ In this limit the Green functions
(\ref{G}) and (\ref{KelvinGo}) for similarly restricted $u$s and $v$s are
given by%
\begin{align}
G\left(  u_{1}=y_{1}/a,v_{1}=\pi/2;u_{2}=y_{2}/a,v_{2}=\pi/2\right)   &
=-\frac{1}{4\pi a}\left\vert y_{1}-y_{2}\right\vert -\frac{1}{4\pi}\ln\left(
1+e^{-2\left\vert y_{1}-y_{2}\right\vert /a}-2e^{-\left\vert y_{1}%
-y_{2}\right\vert /a}\right) \nonumber\\
&  \underset{a\rightarrow\infty}{\sim}-\frac{1}{2\pi}\ln\left(  \left\vert
y_{1}-y_{2}\right\vert /a\right)  +O\left(  \frac{1}{a}\right)  \ ,
\end{align}

\begin{equation}
G_{o}\left(  u_{1}=y_{1}/a,v_{1}=\pi/2;u_{2}=y_{2}/a,v_{2}=\pi/2\right)
\underset{a\rightarrow\infty}{\sim}-\frac{1}{2\pi}\ln\left(  \frac{\left\vert
y_{1}-y_{2}\right\vert }{\left\vert y_{1}-2Y+y_{2}\right\vert }\right)
+O\left(  \frac{1}{a}\right)  \ .
\end{equation}
Finally then,%
\begin{equation}
\lim_{a\rightarrow\infty}G_{o}\left(  u_{1}=y_{1}/a,v_{1}=\pi/2;u_{2}%
=y_{2}/a,v_{2}=\pi/2\right)  =-\frac{1}{2\pi}\ln\left(  \frac{\left\vert
y_{1}-y_{2}\right\vert }{\left\vert y_{1}+y_{2}-2Y\right\vert }\right)  \ .
\end{equation}
But for $y_{1}>Y$ and $y_{2}>Y$, this is precisely the 2D Green function at
field point $\left(  0,y_{1}\right)  $ for a grounded straight line parallel
to the $x$-axis, passing through the point $\left(  x,y\right)  =\left(
0,Y\right)  $, as obtained by placing at the point $\left(  0,2Y-y_{2}\right)
$ a single negative Kelvin point image of a unit point source placed at
position $\left(  0,y_{2}\right)  $. \ Of course, in this straight line limit
where $x_{1}=x_{2}$ the system is translationally invariant with respect to
$x$, so there is no $x$ dependence in the final Green functions.

When $x_{1}\neq x_{2}$ but both are fixed and small, while $a$ becomes
infinite, a similar but slightly more tedious limit calculation gives the 2D
Green function on the grounded half-plane, namely,
\begin{equation}
G_{half-plane}\left(  x_{1},y_{1};x_{2},y_{2}\right)  =-\frac{1}{2\pi}%
\ln\left(  \frac{\sqrt{\left(  x_{1}-x_{2}\right)  ^{2}+\left(  y_{1}%
-y_{2}\right)  ^{2}}}{\sqrt{\left(  x_{1}-x_{2}\right)  ^{2}+\left(
y_{1}+y_{2}-2Y\right)  ^{2}}}\right)  \ .
\end{equation}
Once again, translational invariance with respect to $x$ accounts for the
dependence on only the difference, $x_{1}-x_{2}$. \ We leave the detailed
derivation of $G_{half-plane}$ from $G_{o}$ for the ellipse as an exercise for
the reader.\footnote{Results given in Appendix A may be helpful.}

\subsection*{Discussion}

The standard problems involving a grounded circular ring in 2D
\cite{CurtrightEtAl}\ or grounded spheres in higher dimensions
\cite{AlshalCurtright}\ can also be easily solved using either the Kelvin or
Sommerfeld methods. \ However, there are many problems where the Kelvin method
is very difficult, if not impossible, to implement, but which are directly
solvable by the Sommerfeld method. \ Grounded semi-infinite planes and the
circular disk in 3D Euclidean space provide well-studied examples
\cite{Sommerfeld,Hobson,Waldmann,DavisReitz,DavisReitzAgain,Duffy}.

Beyond these previously solved examples, the grounded ellipsoid in 3D and
hyper-ellipsoids in higher dimensions are difficult problems that should be
more tractable using Sommerfeld images. \ Existing image methods applied to
these problems are quite involved, and usually require detailed properties of
ellipsoidal harmonics \cite{Dassios}. \ In fact, extant treatments of the
exterior 3D Green function problem for grounded ellipsoids use, in addition to
an interior point image, a \emph{continuous} distribution of Kelvin image
charge on the surface of an interior confocal ellipsoid
\cite{DassiosAgain,XueDeng} (also see Sections 7.4.3 and 7.4.4 in
\cite{Dassios}). \ This non-trivial array of image charges results from
requiring that all such charges reside entirely within the physical interior
of the ellipsoid, \emph{without} invoking a second copy of $\mathbb{E}_{3}$.
\ In our opinion, these treatments are tantamount to walking on broken glass
while bare-footed. \ 

In contrast, the Sommerfeld method applied to a grounded ellipsoid embedded in
$N$ Euclidean dimensions only requires a single point image of the point
source, in complete parallel to the grounded 2D ellipse treated here, albeit
at the cost of introducing a second copy of $\mathbb{E}_{N}$. \ Therefore, in
principle the Sommerfeld method should simplify the analysis required to
construct Green functions for such ellipsoids, both conceptually and
practically. \ \bigskip

\noindent\textbf{Acknowledgements} \ It has been our pleasure to reconsider
this elementary subject during the Sommerfeld Sesquicentennial year. \ This
work was supported in part by a University of Miami Cooper Fellowship and by a
Clark Way Harrison Visiting Professorship at Washington University in Saint
Louis.\bigskip

\subsection*{Appendix A: \ Complex variables}

Let%
\begin{equation}
x+iy=a\left(  \cos v\cosh u+i\sin v\sinh u\right)  =a\cosh\left(  u+iv\right)
\ . \tag{A1}%
\end{equation}
That is to say, $u+iv=\pm\operatorname{arccosh}\left(  \frac{x+iy}{a}\right)
+2i\pi k\mid k\in%
\mathbb{Z}
$. \ Choose the $+$ solution with $k=0$ so that%
\begin{equation}
u=\operatorname{Re}\left(  \operatorname{arccosh}\left(  \frac{x+iy}%
{a}\right)  \right)  \ ,\ \ \ v=\operatorname{Im}\left(
\operatorname{arccosh}\left(  \frac{x+iy}{a}\right)  \right)  \ . \tag{A2}%
\end{equation}
Then find%
\begin{align}
r^{2}  &  =x^{2}+y^{2}=a^{2}\left(  \cosh^{2}u\ \cos^{2}v+\sinh^{2}u\ \sin
^{2}v\right) \nonumber\\
&  =\frac{1}{2}~a^{2}\left(  \cosh2u+\cos2v\right)  =a^{2}%
\operatorname{arccosh}\left(  \frac{x+iy}{a}\right)  \operatorname{arccosh}%
\left(  \frac{x-iy}{a}\right)  \ , \tag{A3}%
\end{align}
as well as%
\begin{align}
x^{2}-y^{2}  &  =a^{2}\left(  \cosh^{2}u\ \cos^{2}v-\sinh^{2}u\ \sin
^{2}v\right)  =\frac{1}{2}~a^{2}\left(  \cosh2u\cos2v+1\right)  \ ,\tag{A4}\\
xy  &  =a^{2}\cosh u\ \cos v\ \sinh u\ \sin v=\frac{1}{4}~a^{2}\sinh
2u\sin2v\ . \tag{A5}%
\end{align}
In addition find%
\begin{align}
\sinh^{2}2u  &  =\frac{1}{2}~\cosh4u-\frac{1}{2}\nonumber\\
&  =\frac{2}{a^{4}}\left(  x^{2}+y^{2}\right)  ^{2}+\frac{2}{a^{2}}\left(
y^{2}-x^{2}\right)  +\frac{2}{a^{4}}\left(  x^{2}+y^{2}\right)  \sqrt{\left(
\left(  x-a\right)  ^{2}+y^{2}\right)  \left(  \left(  x+a\right)  ^{2}%
+y^{2}\right)  }\ , \tag{A6}%
\end{align}
along with%
\begin{equation}
v=\arccos\left(  \frac{x/a}{\cosh u}\right)  =\arcsin\left(  \frac{y/a}{\sinh
u}\right)  \ . \tag{A7}%
\end{equation}

\subsection*{Appendix B: \ Circle $\longleftrightarrow$\ ellipse conformal
mapping}

Define a standard ellipse and its fiducial circle by%
\begin{equation}
\frac{x^{2}}{a^{2}}+\frac{y^{2}}{b^{2}}=1\ ,\ \ \ X^{2}+Y^{2}=\left(
\frac{a+b}{2}\right)  ^{2} \tag{B1}%
\end{equation}
Then circles in the complex $Z=X+iY$ plane are mapped to ellipses in the
complex $z=x+iy$ plane, and vice versa, by \cite{Nehari}%
\begin{equation}
z=Z+\frac{c^{2}}{4Z}\ ,\ \ \ c^{2}=a^{2}-b^{2}\ , \tag{B2}\label{map}%
\end{equation}
By definition for any circle in the $Z$ plane, $R^{2}=\left(  X^{2}%
+Y^{2}\right)  =\left\vert Z\right\vert ^{2}$. \ Expressing $R^{2}$ in terms
of $x$ and $y$ as given by the map (\ref{map}) then leads to
\begin{equation}
1=\frac{R^{2}}{\left(  R^{2}+\frac{1}{4}c^{2}\right)  ^{2}}~x^{2}+\frac{R^{2}%
}{\left(  R^{2}-\frac{1}{4}c^{2}\right)  ^{2}}~y^{2} \tag{B3}%
\end{equation}
This is indeed another ellipse, confocal with the standard ellipse, only now
with
\begin{equation}
a^{2}=R^{2}\left(  1+\frac{c^{2}}{4R^{2}}\right)  ^{2}\ ,\ \ \ b^{2}%
=R^{2}\left(  1-\frac{c^{2}}{4R^{2}}\right)  ^{2}\ ,\ \ \ a^{2}-b^{2}=c^{2}\ .
\tag{B4}%
\end{equation}
The point is, concentric circles centered on the origin of the $Z$-plane are
mapped by (\ref{map}) to confocal ellipses centered on the origin of the
$z$-plane, and vice versa. \ Moreover, it is obvious and well-known
\cite{Nehari} that the $Z\rightarrow z$ map actually covers the complex
$z$-plane twice: \ Both the interior and the exterior of the fiducial circle
cover the $z$-plane under the map.

But now consider the well-known electrostatics method to ground a circle by
placing an image charge at a point obtained by inversion of the source
location with respect to that grounded circle.\ \ Where does the conformal map
(\ref{map}) take a point $Z$ after it has been inverted with respect to the
circle of radius $\frac{1}{2}\left(  a+b\right)  $? \ The effect of the
inversion is
\begin{equation}
X\rightarrow\widetilde{X}=\left(  \frac{a+b}{2}\right)  ^{2}\frac{X}{R^{2}%
}\ ,\ \ \ Y\rightarrow\widetilde{Y}=\left(  \frac{a+b}{2}\right)  ^{2}\frac
{Y}{R^{2}}\ . \tag{B5}%
\end{equation}
That is to say,
\begin{equation}
\widetilde{R}^{2}=\widetilde{X}^{2}+\widetilde{Y}^{2}=\left(  \frac{a+b}%
{2}\right)  ^{4}\frac{1}{R^{2}}\ ,\ \ \ \widetilde{Z}=\widetilde{X}%
+i\widetilde{Y}=\frac{1}{R^{2}}\left(  \frac{a+b}{2}\right)  ^{2}Z\ . \tag{B6}%
\end{equation}
So then, the conformal map of this inverted point gives
\begin{equation}
\widetilde{z}=\widetilde{Z}+\frac{c^{2}}{4\widetilde{Z}}=\frac{1}{R^{2}%
}\left(  \frac{a+b}{2}\right)  ^{2}\left(  Z+\frac{c^{2}}{\frac{4}{R^{4}%
}\left(  \frac{a+b}{2}\right)  ^{4}Z}\right)  \tag{B7}%
\end{equation}
For example, suppose $a=3$ and $b=1$, then $\left(  a+b\right)  /2=2$ and
$c^{2}=8$. \ Then%
\begin{align}
x  &  =\left(  1+\frac{c^{2}}{4R^{2}}\right)  X\ ,\ \ \ y=\left(
1-\frac{c^{2}}{4R^{2}}\right)  Y\nonumber\\
\widetilde{x}  &  =\left(  1+\frac{c^{2}}{4\widetilde{R}^{2}}\right)
\widetilde{X}\ ,\ \ \ \widetilde{y}=\left(  1-\frac{c^{2}}{4\widetilde{R}^{2}%
}\right)  \widetilde{Y}\tag{B8}\\
\widetilde{X}  &  =\left(  \frac{a+b}{2}\right)  ^{2}\frac{X}{R^{2}%
}\ ,\ \ \ \widetilde{Y}=\left(  \frac{a+b}{2}\right)  ^{2}\frac{Y}{R^{2}%
}\nonumber
\end{align}
More specifically, consider%
\begin{align}
\left.  \left(  \widetilde{X},\widetilde{Y}\right)  \right\vert
_{a=b=2,X=2.5\cos\theta,Y=2.5\sin\theta}  &  =\left(  2^{2}\times\frac{1}%
{2.5}\cos\theta,2^{2}\times\frac{1}{2.5}\sin\theta\right) \tag{B9}\\
\left.  \left(  \widetilde{x},\widetilde{y}\right)  \right\vert
_{a=b=2,X=2.5\cos\theta,Y=2.5\sin\theta}  &  =\left(  \left(  1+\frac
{8}{4\left(  \frac{2^{2}}{2.5}\right)  ^{2}}\right)  2^{2}\times\frac{1}%
{2.5}\cos\theta,\left(  1-\frac{8}{4\left(  \frac{2^{2}}{2.5}\right)  ^{2}%
}\right)  2^{2}\times\frac{1}{2.5}\sin\theta\right) \nonumber
\end{align}
For other points, see Figures B1 and B2. \ Upon comparing these two Figures,
the various curves are related by the map (\ref{map}). \ Thus the solid or
dashed circles shown in Figure B1 map to the solid or dashed ellipses of the
same color shown in Figure B2, and vice versa. \ Also, the light gray straight
radial line in Figure B1 maps to the light gray hyperbolic curve in Figure B2,
and similarly for other such radial lines.

\hspace{-1in}%
{\parbox[b]{8.0739in}{\begin{center}
\includegraphics[scale=.4]{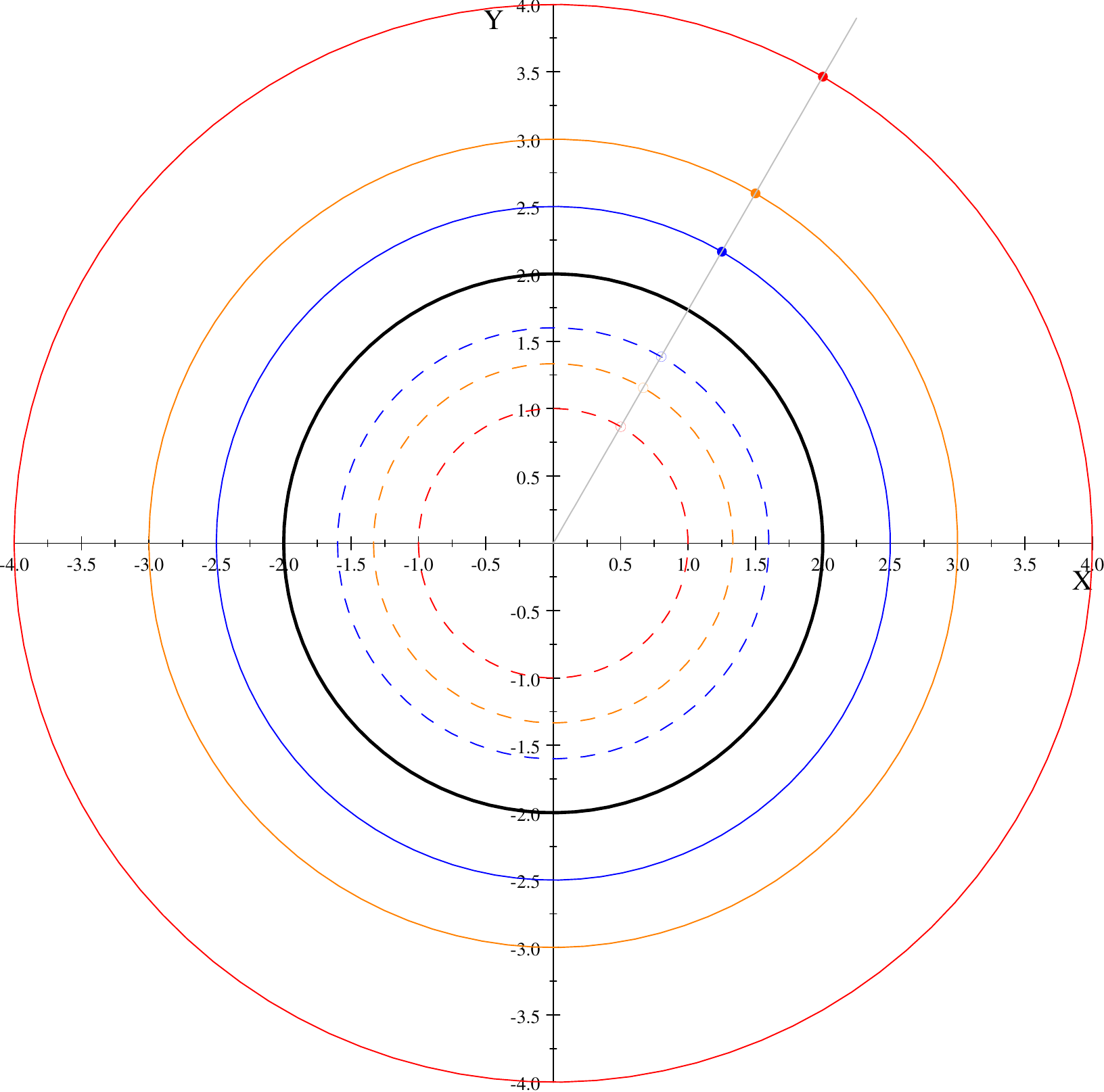}%
\\
Figure B1
\end{center}}}

\hspace{-1in}

\begin{center}
Various source (solid color curves) and Kelvin image (dashed color curves)
charge locations for a grounded circle (shown in black). \ For one-to-one
point source $\leftrightarrow$ point image pairing, only one copy of the plane
is needed.
\end{center}

\hspace{-1in}%
{\parbox[b]{8.0739in}{\begin{center}
\includegraphics[scale=.4]{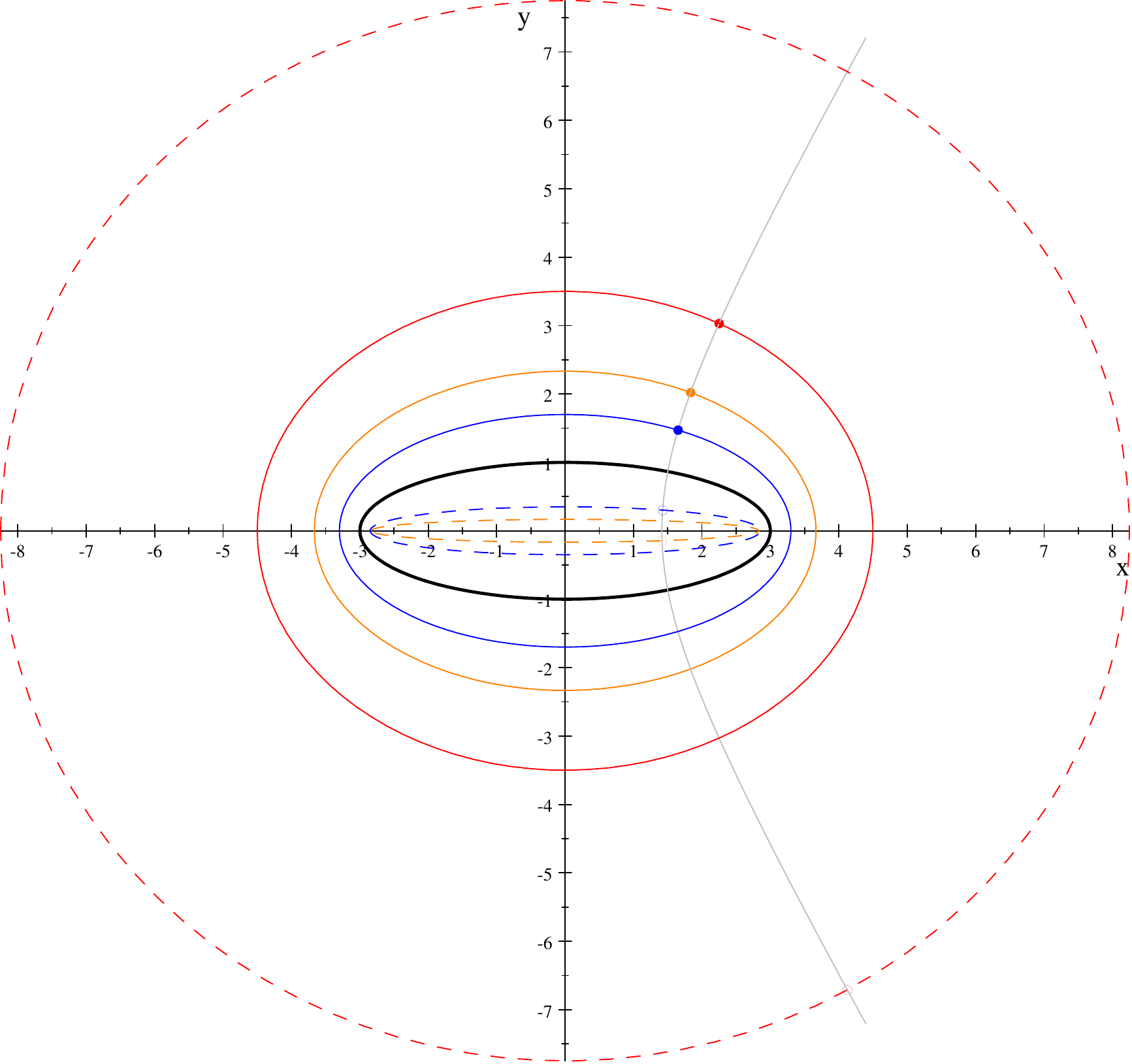}%
\\
Figure B2
\end{center}}}

\begin{center}
Various source (solid color curves) and Kelvin image (dashed color curves)
charge locations for a grounded ellipse (shown in black). \ For one-to-one
point source $\leftrightarrow$ point image pairing, two copies of the plane
are now needed.
\end{center}

\bigskip

These Figures reproduce and confirm the explanation in the text that made use
of real variables, namely, two copies of the plane are required to ground the
ellipse using a single point image for each point source. \ The image
locations shown by the orange and red dashed curves in Figure B2 are actually
on the second branch of the doubled plane.

\newpage

\subsection*{Appendix C: \ Contour plots of $G_{o}$}

Consider the grounded ellipse defined by $\left(  x,y\right)  =\left(
\cos\left(  v\right)  \cosh\left(  1\right)  ,\sin\left(  v\right)
\sinh\left(  1\right)  \right)  $ for $0\leq v\leq2\pi$. \ Three dimensional
contour plots of $G_{o}\geq0$, as functions of the field points on the
$xy$-plane, are shown in the following Figures for three representative point
source locations, with values near the point source truncated at $G_{o}=0.25$.
\ (For an animated version, with source locations varied for $0\leq v\leq2\pi
$, please see
\href{http://www.physics.miami.edu/~curtright/GoRotatingFigure.gif}{this URL}.)

\hspace{-1in}%
\raisebox{-0.0579in}{\parbox[b]{8.0298in}{\begin{center}
\includegraphics{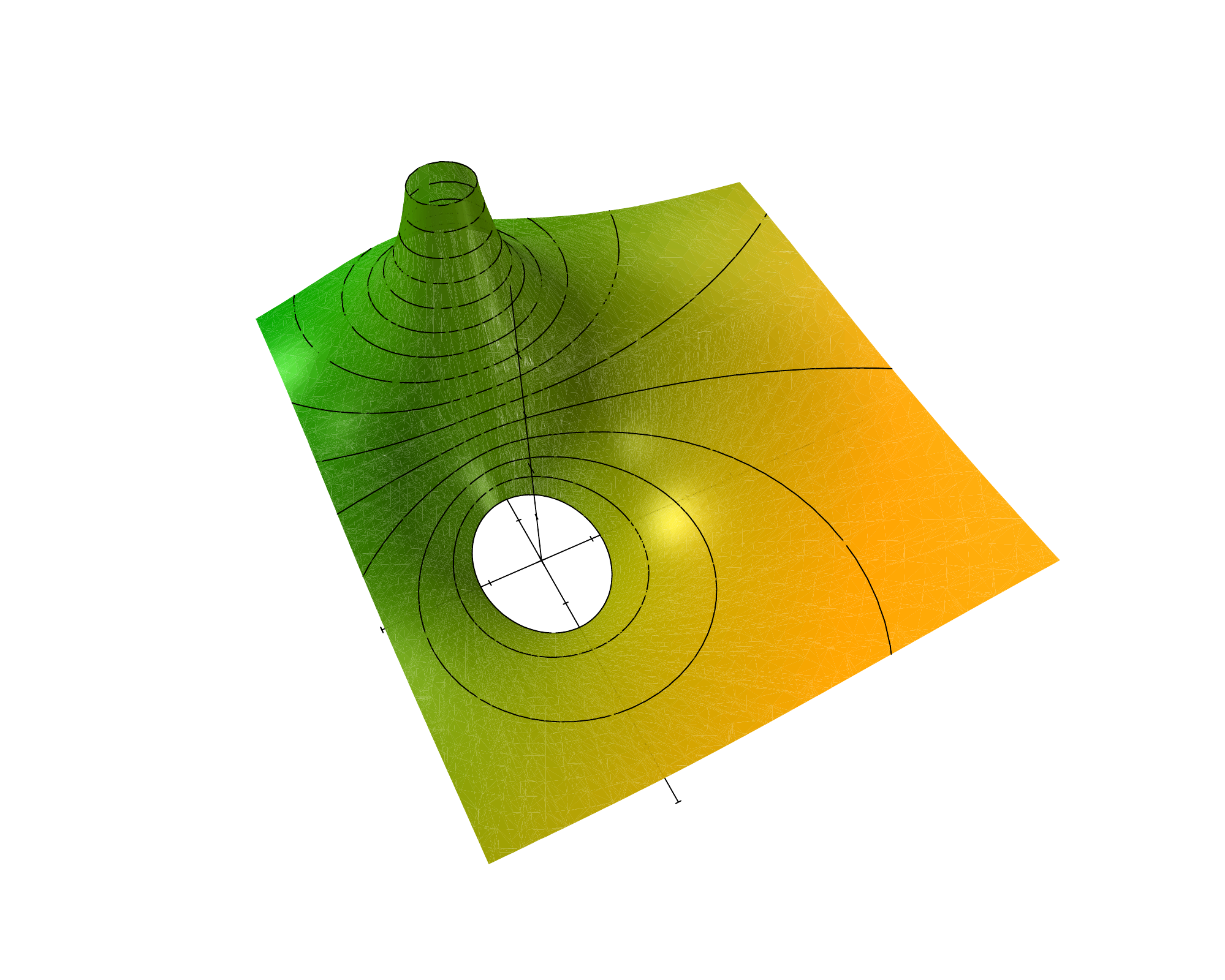}%
\\
Figure C1: \ Contour plot of $G_{o}\left(  x,y\right)  $ with point source at
$\left(  x,y\right)  =\left(  \cos\left(  \pi\right)  \cosh\left(  \frac{3}%
{2}\right)  ,\sin\left(  \pi\right)  \sinh\left(  \frac{3}{2}\right)  \right)
$..
\end{center}}}

\hspace{-1in}%
\raisebox{-0.0579in}{\parbox[b]{8.0298in}{\begin{center}
\includegraphics{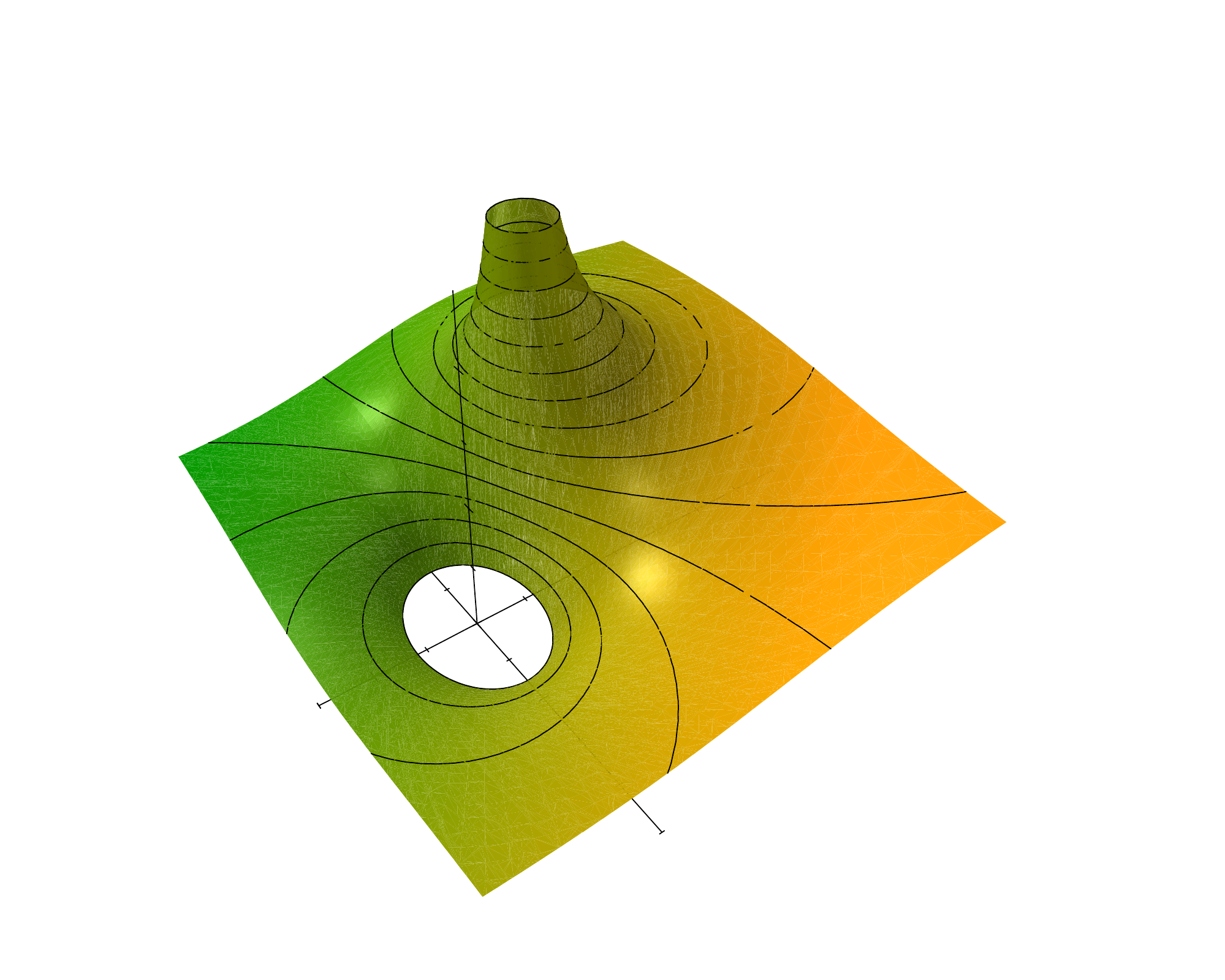}%
\\
Figure C2: \ Contour plot of $G_{o}\left(  x,y\right)  $ with point source at
$\left(  x,y\right)  =\left(  \cos\left(  \frac{2}{3}\pi\right)  \cosh\left(
\frac{3}{2}\right)  ,\sin\left(  \frac{2}{3}\pi\right)  \sinh\left(  \frac
{3}{2}\right)  \right)  $..
\end{center}}}

\hspace{-1in}%
\raisebox{-0.0579in}{\parbox[b]{8.0298in}{\begin{center}
\includegraphics{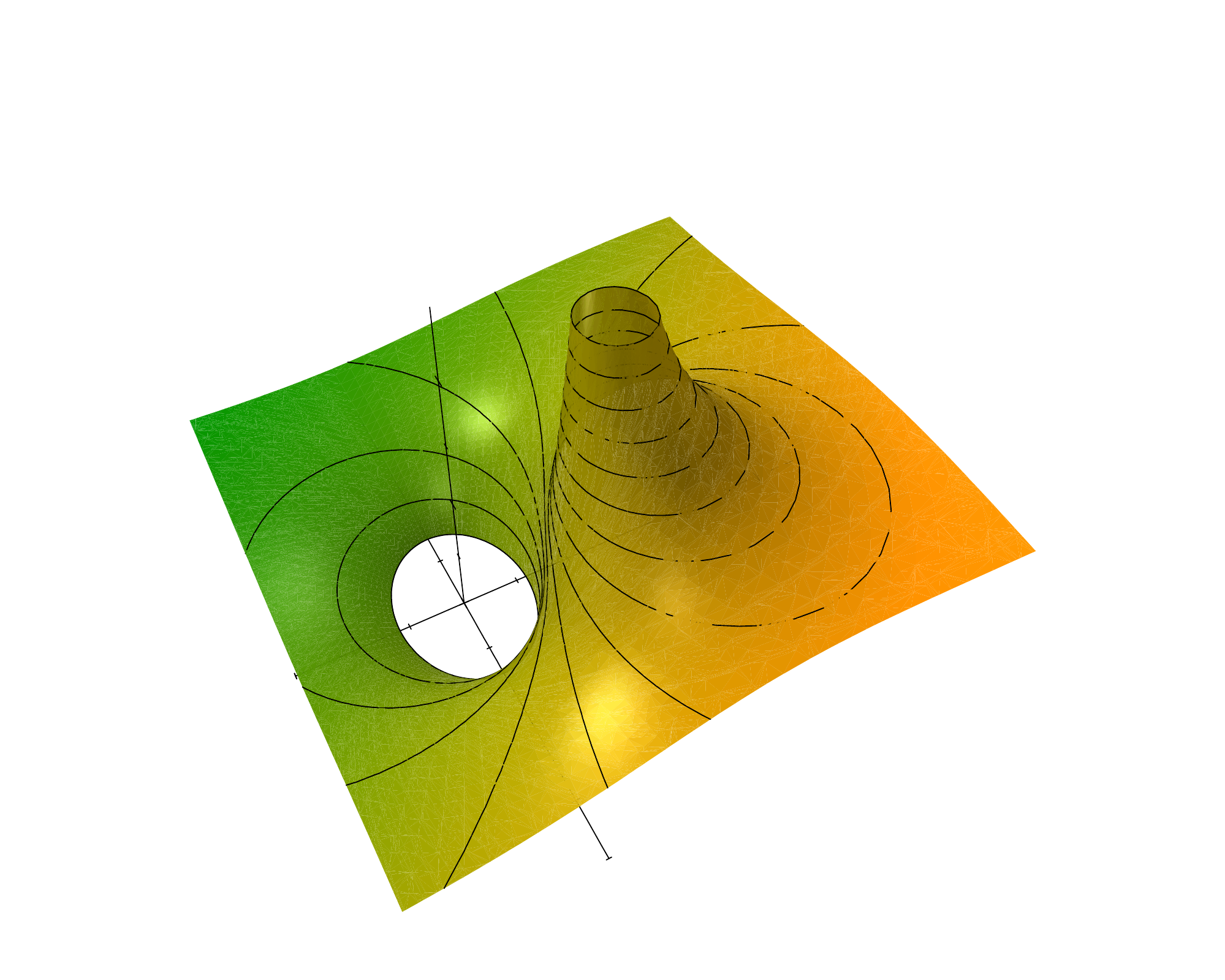}%
\\
Figure C3: \ Contour plot of $G_{o}\left(  x,y\right)  $ with point source at
$\left(  x,y\right)  =\left(  \cos\left(  \frac{1}{3}\pi\right)  \cosh\left(
\frac{3}{2}\right)  ,\sin\left(  \frac{1}{3}\pi\right)  \sinh\left(  \frac
{3}{2}\right)  \right)  $..
\end{center}}}


\newpage

\begin{thebibliography}{99}                                                                                               %


\bibitem {AlshalCurtright}H Alshal and T Curtright, \textquotedblleft Grounded
Hyperspheres as Squashed Wormholes\textquotedblright%
\ \href{https://arxiv.org/abs/1806.03762}{arXiv:1806.03762 [physics.class-ph]}.

\bibitem {CurtrightEtAl}T Curtright, H Alshal, P Baral, S Huang, J Liu, K
Tamang, X Zhang, and Y Zhang. \textquotedblleft The Conducting Ring Viewed as
a Wormhole\textquotedblright%
\ \href{https://arxiv.org/abs/1805.11147}{arXiv:1805.11147 [physics.class-ph]}.

\bibitem {Dassios}G Dassios,
\textit{\href{https://www.amazon.com/Ellipsoidal-Harmonics-Encyclopedia-Mathematics-Applications-ebook/dp/B00D2WQARA/ref=sr_1_1?s=digital-text&ie=UTF8&qid=1533862975&sr=1-1}{\textit{Ellipsoidal
Harmonics}}}, Cambridge University Press (2013) ISBN-13: 978-0521113090.

\bibitem {DassiosAgain}G Dassios, \textquotedblleft Directional dependent
Green's function and Kelvin images\textquotedblright%
\ \href{https://doi.org/10.1007/s00419-012-0669-6}{Arch. Appl. Mech. 82 (2012)
1325--1335}.

\bibitem {DavisReitz}L C Davis and J R Reitz, \textquotedblleft Solution to
potential problems near a conducting semi-infinite sheet or conducting
disc\textquotedblright\ \href{https://doi.org/10.1119/1.1976616}{Am. J. Phys.
39 (1971) 1255-1265}.

\bibitem {DavisReitzAgain}L C Davis and J R Reitz, \textquotedblleft Solution
of potential problems near the corner of a conductor\textquotedblright%
\ \href{https://doi.org/10.1063/1.522671}{J. Math. Phys. 16 (1975) 1219--1226}.

\bibitem {Duffy}D G Duffy,
\textit{\href{https://www.crcpress.com/Greens-Functions-with-Applications-Second-Edition/Duffy/p/book/9781138894464}{\textit{Green's
Functions with Applications}}}, Second Edition, CRC Press (2017) ISBN-13: 978-1482251029.

\bibitem {Eckert}M Eckert,
\textit{\href{https://www.amazon.com/Arnold-Sommerfeld-Science-Turbulent-1868-1951/dp/1461474604/ref=sr_1_7?s=books&ie=UTF8&qid=1524241334&sr=1-7&keywords=sommerfeld}{\textit{Arnold
Sommerfeld: \ Science, Life and Turbulent Times 1868-1951}}}, Springer-Verlag
(2013) ISBN-13: 978-1461474609.

\bibitem {Green}G Green,
\href{http://en.wikipedia.org/wiki/An_Essay_on_the_Application_of_Mathematical_Analysis_to_the_Theories_of_Electricity_and_Magnetism}{\textit{An
Essay}} \textit{on the Application of Mathematical Analysis to the Theories of
Electricity and Magnetism}, Nottingham (1828).

\bibitem {Hobson}E W Hobson, \textquotedblleft On Green's function for a
circular disc, with application to electrostatic problems\textquotedblright%
\ \href{https://babel.hathitrust.org/cgi/pt?id=coo.31924069328965;view=1up;seq=315}{Trans.
Cambridge Philos. Soc. 18 (1900) 277- 291}.

\bibitem {Nehari}Z Nehari,
\textit{\href{https://www.amazon.com/Conformal-Mapping-Dover-Books-Mathematics-ebook/dp/B00A73A348/ref=sr_1_fkmr0_1?s=digital-text&ie=UTF8&qid=1542954089&sr=1-1-fkmr0&keywords=conformal+transformations+nehari}{\textit{Conformal
Mapping}}, }Dover Publications (2011), especially pp 269-271, Eqn(4) et seq.

\bibitem {Sommerfeld}A Sommerfeld, \textquotedblleft\"{U}ber verzweigte
Potentiale im Raum\textquotedblright%
\ \href{https://doi.org/10.1112/plms/s1-28.1.395}{Proc. London Math. Soc.
(1896) s1-28 (1): 395-429}; ibid. 30 (1899) 161.

\bibitem {KelvinTait}W Thomson and P G Tait, \textit{Treatise on Natural
Philosophy},
\href{http://babel.hathitrust.org/cgi/pt?id=wu.89068226117;view=1up;seq=13}{Part
I} \&
\href{http://babel.hathitrust.org/cgi/pt?id=wu.89068225952;view=1up;seq=5}{Part
II}, Cambridge University Press (1879 \& 1883).

\bibitem {Waldmann}L Waldmann, \textquotedblleft Zwei Anwendungen der
Sommerfeld'schen Methode der verzweigten Potentiale\textquotedblright%
\ \href{https://babel.hathitrust.org/cgi/pt/search?id=mdp.39015076063125;view=1up;seq=5;q1=Waldmann;start=1;sz=10;page=search;orient=0}{Physikalische
Zeitscrift 38 (1937) 654--663}

\bibitem {XueDeng}C Xue and S Deng, \textquotedblleft Green's function and
image system for the Laplace operator in the prolate spheroidal
geometry\textquotedblright\ \href{https://doi.org/10.1063/1.4974156}{AIP
Advances 7 (2017) 015024}.
\end{thebibliography}
\end{document}